\documentstyle[aps,twocolumn,psfig]{revtex}
\begin{document}%************************************************************
\title{Coulomb correlations and coherent charge tunneling in 
mesoscopic coupled rings}
\author{C.~M.~Canali$^1$, W.~Stephan$^2$, L.~Y.~Gorelik$^1$,
R.~I.~Shekhter$^1$ and M.~Jonson$^1$}
\address{$^1$Dept. of Applied Physics, Chalmers University of Technology and
G\"oteborg University, S-412 96 G\"oteborg, Sweden\\
$^2$Max-Planck-Institut f\"ur Physik komplexer Systeme,
Bayreuther Str. 40, D-01187 Dresden, Germany
\\[3pt]
(1 October 1997)%\today)
\\ \medskip}
\author{\small\parbox{14cm}
{\small
We study the effect of a strong
electron-electron (e-e) interaction
in a system of two concentric one-dimensional rings with
incommensurate areas $A_1$ and $A_2$,
coupled by a tunnel amplitude.
For noninteracting particles the magnetic moment (persistent current) $m$ of
the many-body ground state and
first excited states is an irregular function of the external
magnetic field. 
In contrast, we show that when strong e-e
interactions are present
the magnetic field dependence of $m$
becomes periodic.
In such a strongly correlated system disorder can only be caused by
inter-ring charge fluctuations, controllable by a gate voltage. 
The oscillation period of $m$ is
$ \propto 1/(A_1 + A_2)$ if fluctuations are
suppressed. Coherent inter-ring tunneling
doubles the period when charge fluctuations are
allowed.
%The amplitude of the magnetic moment and the parity effects
%are also modified by interactions.
%
\\[3pt]PACS numbers: 71.27.+a, 73.20.Dx, 72.15.Nj}}
\address{}\maketitle
%\bibliographystyle{simpl1}
%\narrowtext
\eject
Persistent currents in quasi onedimensional 
mesoscopic rings\cite{buttiker}
are a phase sensitive probe of the conductor equilibrium properties.
The intriguing role played by e-e interactions
in such a phase coherent phenomenon
continues to be the source
of intensive investigation. On one hand theoretical work\cite{loss}
has shown that interaction
does not influence the persistent currents in clean single-channel
rings, in agreement with experiments on a single ballistic 
semiconductor
ring with few channels\cite{pcexp1ch}.
On the other hand many other theoretical studies\cite{theory} 
have explored the possibility of linking the e-e interaction to 
the puzzlingly large current amplitudes found experimentally
in disordered metallic (i.e. many-channel) rings\cite{pcexpmch}.
Such amplitudes cannot be accounted for within a single-particle theory.
No general consensus has been found yet
on the effects of interaction in these systems.

As an example where e-e interaction
is expected to play a distinctive role and cause interesting effects,
we consider here a system of
two concentric ballistic rings, coupled by a tunnel amplitude. 
The areas enclosed by the two rings are
$A_1$ and $A_2$ respectively, and a perpendicular magnetic
field $B$ is present. 
Our aim is to investigate how strong interactions
affect the many-body ground-state and the first excited states as
a function of the applied magnetic field. We will show that 
the interaction creates a spatially ordered rigid structure, where
the two rings are tied together and
charge fluctuations tend to be suppressed. 
Consequently the flux dependence of the lowest
energy levels becomes periodic with period $\propto 1/(A_1 + A_2)$, 
in contrast to the irregular aperiodic
behavior of the free-particle case. By means of a gate voltage
one can induce the coherent transfer of a single electronic charge
from one ring to the other. As a result of the 
coherent charge fluctuations between the rings, the oscillation period
of the lowest levels is doubled.
Charge transfer induced effects in different mesoscopic samples have been
recently studied by B\"uttiker and Stafford\cite{BandS}.

We choose a lattice model of two concentric one-dimensional rings 
with ${\cal N}$ sites per ring (see Fig.~1),
described by the following second quantized Hamiltonian
\begin{eqnarray}
{\cal H} = &&
-t\sum_{i,p}
e^{i2\pi \varphi_p/{\cal N}}\,c^{\dagger}_{i,p}
c^{\phantom\dagger}_{i+1,p}
%\nonumber\\&&
-\sum_{i}t_{12}(i)c^{\dagger}_{i,1}c^{\phantom\dagger}_{i,2}+\; H.c.
\nonumber\\&&
 + \sum_{i,j,p,p'} V(|{\bf r}_{i,p} - {\bf r}_{j,p'}|)\,n_{i,p}n_{j,p'}
\end{eqnarray}
The operators $c^{\dagger}_{i,p}$ and $c^{\phantom\dagger}_{i,p}$
create and destroy a spinless electron at site $i$ of ring $p$.
($i,j=1,\dots,{\cal N}$ are site indexes 
and $p, p' = 1,2$ are ring indexes).
The term proportional to $t$ describes the {\it intra-ring} hopping,
affected by the magnetic field through the 
phases
$\varphi_{p} =  B A_{p}/\phi_0$,
where $\phi_0 = c h/e$ is the flux quantum.
The site-dependent amplitude $t_{12}(i)$
represents {\it inter-ring} tunneling.
The last term is the 
translationally invariant e-e interaction. 
The number of electrons in each ring, ${N}_p$, is
not a good quantum number if the rings are coupled.
The expectation values $\langle{N}_p\rangle$ will however satisfy 
the constraint $\langle {N}_1\rangle +\langle  {N}_2\rangle= {N}$;
${N}$ is the total number of electrons in the system.%\cite{note1}.

We first discuss the case of two decoupled [$t_{12}(i)=0$]
noninteracting rings.
The single-particle spectrum of each
ring is given by 
$\epsilon_p(k_i) = -2 t\cos[(k_i + 2\pi \varphi_p/{\cal N})] =
-2 t\cos[(k_i + 2\pi H\,a_p/{\cal N})]$, with 
$k_i =2\pi\, i/{\cal N},\  i=0,\pm 1,\dots \pm {\cal N}/2$.
Here we have introduced two independent, dimensionless parameters,  
$H$ (magnetic field) and 
$a_p$ (area of ring $p$). 
When $H=1$ and $a_p=1$ the flux enclosed by ring $p$ is $\phi_0$.
%The ground-state
%energy of the system $E_0(H)$ is obtained by filling the lowest ${N}$
%energy levels taken from the spectra of both rings. 
As $H$ varies
the levels move up and down; at some values
of $H$ they can cross. As a result, the particle at the
Fermi energy can switch to an empty state of different ``velocity'',
causing a cusp in the magnetic field dependence of the 
ground state energy, $E_0(H)$.
Cusps in $E_0(H)$ will cause jumps in the total magnetic moment of the
ground state, $m(H) \equiv {\partial E_0(H) / \partial H}$.
When ${N}$ is even one can see that, in the ground-state,
each ring of the system contains always ${N}/2$ particles.
When flux-induced level crossing occurs, the particle at the Fermi energy
will change place to an empty state of the {\it same} ring with opposite
velocity. 
Therefore $m(H)$ will consist of the
superposition of the contributions of two independent rings,
each one oscillating with its own period $1/a_p$ as a function of $H$.
If the areas of the two rings are incommensurate,
-- throughout this work we have considered the values
$a_1 = [\sqrt(5)-1]/2\approx 0.618$ and $a_2=1.5$ --
the resulting
total $m(H)$ is a bound but {\it aperiodic} function of $H$, with sharp jumps
occurring at multiple of $1/a_p,\  p=1,2$.
On the other hand
when ${N}$ is odd
the relevant level crossings
involve always two states belonging to {\it different} rings: an occupied
state of the more populated ring is depleted in favor of an empty
state of the other ring. In this case
the aperiodic $m(H)$ contains
jumps at values of $H$ which are multiples of $1/(a_1 \pm a_2)$,
corresponding to level crossings between states of different rings
moving in opposite or equal direction respectively.

To first order in perturbation theory a small tunneling amplitude  
$|t_{12}(i)|<< \Delta = 8\, t/{\cal N}$ 
(here $\Delta$ is the mean level spacing for each single ring) 
will couple states of different rings. 
If $t_{12}(i)$ is also site independent,
the spectrum will hardly be modified and, in particular,
the jumps in $m(H)$ will remain sharp, since they always involve
crossing between levels of different momenta which cannot be coupled
by a homogeneous hopping. However, if $t_{12}(i)$ is not
homogeneous, states of different momenta can be coupled, causing
a rounding of the sharp jumps in $m(H)$. 
If $|t_{12}(i)| \sim \Delta$
the energy levels of the two rings will be strongly hybridized
and $m(H)$ will depend on the specific form of
$t_{12}(i)$. 

In order to see how strong e-e interactions modify this scenario
we have resorted to exact
numerical calculations on small systems, using the Lanczos algorithm
to compute the many-body ground state and the first low-lying excitations.
We have considered systems with ${\cal N}=11$ sites per ring and
${N} =4,5,6,7,8$ particles.
We will first discuss the case of a short range
interaction, coupling two nearest-neighbor
sites $(i,p)$ and $(i+1,p)$
in the same ring with matrix element $V_{11}$,
and opposite sites $(i,1)$ and $(i,2)$ in different rings
with matrix element $V_{12}$.
The relevant energy parameter, indicating the relative strength
of the interaction, is $V_{p p'}/t$.
Plots of the magnetic moment for systems of $N= 6,7,8$ particles
interacting with
$V_{11} =6.5\,t,\ \ V_{12} = 6.3\,t$
are shown in Fig.~2.
The effect of the interaction is
visibly spectacular, particularly when ${N}$ is even:
the aperiodic $H$-dependence of $m(H)$ of the
%FIG2
noninteracting case is replaced 
by a perfectly {\it periodic} function, oscillating with period
$H_0 = 1/(a_1 + a_2)$, exactly as we would have for a single ring with area 
$a_{\rm tot}= (a_1 + a_2)$. 
The periodic $m(H)$ develops smoothly upon
increasing the interaction and once it is reached it becomes
independent of the interaction strength.
The amplitude of $m(H)$ in the strongly interacting regime
is approximately one half of the amplitude of the noninteracting case.
When $N$ is even, the persistent current 
$I(H) \equiv -c\, m(H)$ is diamagnetic, 
opposite to the single-ring case.
As shown in Fig.~2(c),
an inter-ring hopping $t_{12}(i)$ of any kind 
does not spoil this result, even when its
magnitude is comparable to the intra-ring hopping $t$.
Morover, not only is the ground-state energy
a periodic function of $H$, but so are the first low-lying excited states.
In particular, the low-energy sector of the many-body
spectrum is qualitatively identical to the single-particle spectrum in one ring.
We can also evaluate the particle
number in each ring for each many-body state
by computing the corresponding wavefunction.
The calculation shows that in the strongly interacting regime 
all periodic eigenstates have
$ N_1 = N_2 = {N}/2$
for any value of $H$ when the rings are decoupled.%\cite{note1}. 
Inter-ring charge redistributions, causing $H$-dependent deviations from
the $(N_1= {N}/2,N_2= {N}/2)$ configuration, are suppressed here.
Charge fluctuations occur however in the higher
aperiodic states; these are separated from the ordered low-energy 
sector by an energy gap
that scales with $V_{p p'}/t$. A non-zero
$t_{12}(i) \sim \Delta <<V_{p p'} $ cannot connect these two
sectors of the spectrum. Therefore the low-lying states remain periodic
with $\langle N_1\rangle \approx\langle N_2\rangle \approx {N}/2$, even
when the rings are coupled.

Strong intra-ring interaction is known to generate Wigner-crystal-like
ground states in a single ring\cite{shultz}.
It has also been shown that inter-ring interaction causes a nondissipative
Coulomb drag\cite{rojo} that affects qualitatively the persistent 
currents in each ring\cite{ulloa}.
Our exact results suggest that strong inter-ring interaction
glues together the two stiff Wigner crystals created by the intra-ring
interaction, resulting in a 
rigid structure that rotates as a single solid body
under the effect of the magnetic field.
%FIG2
%\begin{figure}
%\vspace{0.5truecm}
%\epsfxsize=4.0truecm \epsffile{fig2.eps}
%\vspace{0.5truecm}
%%\hspace{0.5truecm}
%\caption{Energy of the ground state and the first three excited
%states vs $H$ for the two-ring system of Fig.~1.
%%The total number of particles is $N = 8$.
%The rings are coupled by the same $t_{12}(i)$.
%(a) Interacting case
%with a short range force, $V_{11} =6.5t$, $V_{12} =6.3t$.
%(b) Noninteracting case.}
%\label{fig2}
%\end{figure}
%
Based on this picture, we can easily see why the
oscillation period of the total magnetic moment of such a
structure is $1/(a_1 + a_2)$.
Indeed the Hamiltonian 
of the two frozen crystals
glued together,
each containing $N/2$ particles,
is found to be
\begin{equation}
\label{ham2wc}
%H & = & {1\over2\,I}
%\Bigl[{\cal P}_{\theta} - {e\over c} {N \over 2} {B \over 2\pi}
%(A_1 + A_2)\Bigr]\nonumber\\
{\cal H} 
= {\hbar^2\over 2\,{\cal I}} 
\Bigl[-i\,{\partial\over\partial\theta} - {N \over 2} 
H {(a_1 + a_2)}\Bigr]^2
\end{equation}
where $-i\,\hbar {\partial \over \partial \theta}$ 
is the canonical momentum conjugated to $\theta$,
the rotation angle of the solid around an axis passing through
the center of the rings; ${\cal I} = m_e\, {N \over 2\pi} (A_1 + A_2)$ 
is the moment of inertia. In zero flux, the eigenstates
$\Psi_n(\theta) =\exp (i\,n\,\theta)$ must
satisfy the boundary condition 
$\Psi_n(\theta) = \Psi_n\bigl(\theta + 2\pi/(N/2)\bigr)$ due to the ordered
crystal structure. This immediately implies the announced $1/(a_1 + a_2)$
periodicity as a function of $H$.

The $m(H)$ for odd ${N}$ [see Fig~2(b)]
becomes also periodic in presence of a
strong interaction,
with the same period $H_0$. One significant
difference with respect to even ${N}$ is the shift of $m(H)$ by $H_0/2$:
the system is now paramagnetic.
%Furthermore the calculation of the ring occupancy shows that
%there are now two electronic configurations,
%$\bigl (N_1 = {N \pm 1\over 2},\,  N_2 = {N \mp 1\over 2}\bigr )$,
%corresponding to two quasi-degenerate states,
%which alternate in the ground state as a function of $H$.
%The glitches in $m(H)$ seen in Fig.~2(b)
%correspond to switches between these quasi-degenerate states,
%occurring with period $1/(a_1 - a_2)$.

We have also studied a more realistic unscreened
long range interaction coupling all the sites with matrix element
$V(|{\bf r}_{i,p} - {\bf r}_{j,p'}|) = V_c R_1/|{\bf r}_{i,p} -{\bf r}_{j,p'}|= 
V_c\,\sqrt{a_1}/ \Bigl [a_p + a_{p'} -2\sqrt{a_p\; a_{p'}} 
\cos \bigl(2\pi(i-j)/{\cal N}\bigr)\Bigr ]^{1/2}$.
The dimensionless interaction constant is 
$\alpha= V_c R_1/\hbar v_{\rm F} = (V_c/t)[{\cal N}/ 4\pi \sin (k_{\rm F})]$.
In the strong interaction limit ($\alpha >5$) 
the Hamiltonian Eq.~[1] 
has an unphysical ground state where all the particles are located
in the outer ring. Obviously a realistic Hamiltonian should contain
the contribution of a positive background preventing the depletion
of the inner ring. This is usually done by replacing
$n_{i,p}$ by $n_{i,p} -K$ in Eq.~[1], where $K$ represents the average
positive jellium which equals the average electron density.
Such a replacement is formally equivalent 
to add in the Hamiltonian Eq.~[1] a potential
difference between the rings, 
$H_{g} = \sum_{i,p} \epsilon_{p}n_{i,p}$.
Since we also want to consider the effect of a real external gate voltage 
$V_g = \epsilon_1 -  \epsilon_2$ we will assume that the 
background is incorporated in $V_g$, which we will take as a parameter.
Typically a voltage
difference $V_g \approx -V_c/2$ ensures
an almost equal population in the two rings.
In Fig.~3 we plot $m(H)$ for $N=5, 6$ particles interacting
with a Coulomb force of intensity $V_c = 10\, t$, for two values of
$V_g$.
Let us consider first $N=6$ and $t_{12}=0$ (black line
in Fig.~3(a)(b)).
For $V_g =-0.75\,V_c$, $m(H)$ is perfectly periodic 
with the expected period $1/(a_1 + a_2)$.
On the other hand imperfections in the periodic pattern appear
for $V_g =-0.5\,V_c$.
Note in particular the small glitches at $H=0.3,1$
and the splitting of the
peak at $H=3$ in Fig.~3(a).
In fact the calculation of the ring occupancy shows
that while for $V_g =-0.75\,V_c$ the electronic configuration
$(N_1 =3, N_2=3)$ is stable for all 
values of H, when $V_g =0.5\,V_c$ the
two configurations $(N_1 =3, N_2=3)$ and $(N_1 =2, N_2=4)$ alternate in the
ground state. 
The aperiodic jumps in Fig.~3(a) correspond to a switch from
one configuration to another as $H$ varies. 
It is instructive to look at the $H$ dependence of the 
lowest four energy levels, plotted in Fig.~4, and their corresponding
ring occupancy.
Fig.~4(a) shows that
when $V_g= -0.5\,V_c$ these four levels are often two by two degenerate,
the degeneracy being caused
by the presence of the two electronic configurations 
$(N_1 =3, N_2=3)$ and $(N_1 =2, N_2=4)$
mentioned above. 
$V_g$ is not yet strong enough to enforce a fixed electronic 
configuration in the two rings. 
As a result the levels are not completely
periodic and they are sensitive to inter-ring tunneling-- see grey line of
Fig~3(a). However if we increase $V_g$ up
to $0.75\,V_c$, the configuration $(N_1 =3, N_2=3)$ is locked
in the first three states. 
Charge redistribution becomes possible only starting 
from the fourth level,
with the appearance of the
$(N_1 =2, N_2=4)$ configuration at some values of $H$.
The important consequence is the 
ordering of the first three levels, which is robust against 
tunneling between the rings, as shown in Fig.~3(b) and
Fig.~4(c). 

%FIG3
%FIG4
The example of 6 particles is representative of what should happen
in a real mesoscopic system with $N_1\sim N_2 \sim 50$, where gate-
and flux-induced charge redistribution between the rings amounts to
small changes in their electron concentration.
If the gate voltage
enforces an almost equal population
and prevents charge fluctuations between the rings as $H$ varies,
then the e-e Coulomb interaction
can fully develop the ordering effect.
%Disorder in this strongly correlated system can only be induced by
%charge redistribution, which is controllable
%by tuning $V_g$.  

The presence, at some values of  $V_g$, of two quasi-degenerate states
that differ by one single electron being moved
from one ring to the other gives rise to the possibility of coherent
tunneling\cite{BandS} in presence of strong correlation. 
This effect is best illustrated for odd $N$ when the swapping of
one electron between the rings generates the two configurations
$\bigl (N_1 = {N \pm 1\over 2},\,  N_2 = {N \mp 1\over 2}\bigr )$. 
In this case a homogeneous $t_{12}$ coupling the two rings allows
the swapping electron to occupy an hybridized orbital 
involving two opposite sites of the two rings with equal probability.
Since the particles are indistinguishable
all of them can be coherently involved in this process.
This means that all the $N$ particles will spend half of their time
on either ring, that is 
$\langle N_1\rangle \approx \langle N_2\rangle \approx N/2$, 
which is equivalent to having $N$ particles on
a single ring of area $(a_1 + a_2)/2$. The period of the persistent
current of such ``dynamically-glued'' Wigner crystals will be 
$2\,H_0 = 2/(a_1 + a_2)$, twice as large
as the previous case, where charge fluctuations were totally suppressed.
The numerical results for $N=5$ shown in Fig.~3(c) confirms fully
this prediction: the presence of a homogeneous $t_{12}$ coupling
coherently the quasi-degenerate configurations 
$(N_1=2, N_2=3)$ and $(N_1=2, N_2=3)$
doubles the oscillation period of $m(H)$. Furthermore also the {\it amplitude}
of $m(H)$ is factor of 2 larger than when $t_{12}=0$. 
A similar result
holds approximately for $N=6$, see Fig.~3(b), 
but now the oscillating period must be generalized to 
$N/ ( \langle N_1\rangle a_1 +\langle N_2\rangle a_2)$. 
For $N=6$ the computed averaged occupations are
$\langle N_1\rangle \approx 2.5$, $\langle N_2\rangle \approx 3.5$
and the resulting period is a little smaller than $2/(a_1 + a_2)$.
When $N >>1$ we have
$\langle N_1\rangle/N \approx \langle N_2\rangle/N \approx 1/2$
and we obtain the result that holds exactly for $N=5$.

Our calculations show that spectrum ordering is already well developed
for $\alpha \approx 5$. Experimentally, GaAs-Al/GaAs heterostructures 
with $\alpha > 1$ are available.
Thus the effects that we have found might be seen
in a system of two semiconductor rings. 
Our results might also be connected to
the halving of the oscillation
period of the Aharonov-Bohm effect in Quantum Hall edge states around a
semiconductor antidot\cite{ford}. The experiment in Ref.\cite{ford} 
showed that
when just two edge states encircled the antidot
at approximately the same radius, 
two sets of AB oscillations seemed to lock together, but out
of phase, giving an apparent $(h/e)/2$ oscillation. This seems to correspond
to a double-ring system with $A_1 = A_2$, with periodicity 
proportional to $1/2 A_1$.

In conclusion, we have shown that e-e interaction
orders the flux dependence of the low-energy spectrum 
of a two-ring system.
When strong e-e correlations are present,
disorder can only be induced by inter-ring 
charge redistribution controllable
through an external gate.

We thank I.~V.~Krive for useful discussions. 
Financial support from NFR and TFR is
gratefully acknowledged.

%\acknowledgments

%FIG1
\begin{figure}
\hbox{\psfig{figure=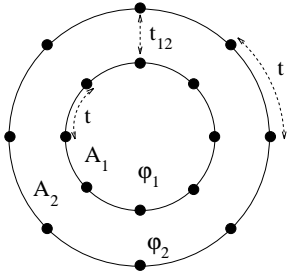,width=6.5cm}}
\caption{Schematic picture of two concentric rings with enclosed areas
$A_1$ and $A_2$. The black dots are the lattice sites.}
\label{fig1}
\end{figure}
%FIG1

%FIG2
\begin{figure}
\vspace{0.5cm}
\hbox{\psfig{figure=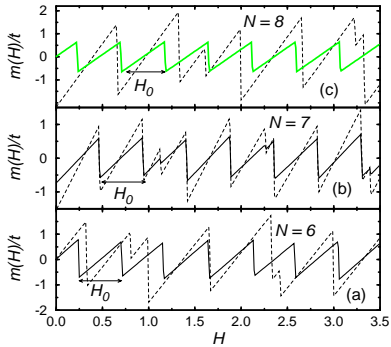,width=6.5cm,angle=-90}}
\vspace{-0.5cm}
\caption{Magnetic moment vs $H$
for the sample of Fig.~1 with ${\cal N} =11$ sites per ring.
The normalized areas are $a_1= [\protect \sqrt(5)-1]/2\approx 0.618$
and $a_2 = 1.5$ respectively. The total number of particles is:
(a)~ ${N}=6$,
(b)~ ${N}=7$ and (c)~ ${N}=8$.
{\it Aperiodic} dashed curves:
decoupled rings [$t_{12}(i)=0$],
noninteracting particles. {\it Periodic} black lines:
$t_{12}(i)=0$,
particles interacting via a short range force;
the oscillation period,
is  $H_0 = 1/(a_1 + a_2)\approx 0.47$.
{\it Periodic} gray line in (c):
$t_{12}(i) = -t/2*\cos ^2(3\pi i/2{\cal N})$,
interacting particles.}
\label{fig2}
\end{figure}
%FIG2

%FIG3
\begin{figure}
\vspace{1cm}
\hbox{\psfig{figure=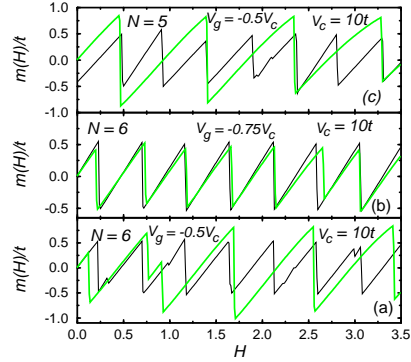,width=6.5cm,angle=-90}}
\vspace{-1cm}
\caption{Magnetic moment vs $H$
for particles interacting with a
Coulomb potential of intensity $V_c = 10\,t$.
$V_g$ is the voltage between the rings.
Black lines:
$t_{12}(i)= 0$;
Gray lines:
$t_{12}(i) = -t/2$.
(a) $N=6$ particles, $V_g = -0.5\, V_c$. (b) $N=6$,
$V_g = -0.75\, V_c$.
(c) $N=5$, $V_g = -0.5\, V_c$.}
\label{fig3}
\end{figure}

%FIG4
\begin{figure}
\vspace{1cm}
\hbox{\psfig{figure=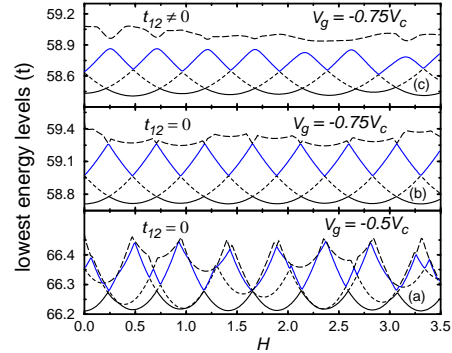,width=6.5cm,angle=-90}}
\vspace{-1cm}
\caption{Energy vs $H$ of the ground state and the first three excited states
for a system of $N=6$ particles, interacting with a Coulomb potential of
intensity $V_c =10t$.
$V_g$ is the voltage between the rings,
(a) $t_{12}=0$, $V_g = 0.5 V_c$.
(b) $t_{12}=0$, $V_g= -0.75\, V_c$.
(c) $t_{12}(i) = -t/2$, $V_g= -0.75\, V_c$.}    %\,\cos^2(2\pi i/{\cal N})$,
\label{fig4}
\end{figure}

\end{document}